\documentclass{ieeetran}
\usepackage{cite}
\usepackage{amsmath,amssymb,amsfonts}
\usepackage{algorithmic}
\usepackage{graphicx}
\usepackage{textcomp}
\usepackage{multirow}
\usepackage{soul}
\usepackage[caption=false,font=footnotesize,subrefformat=parens,labelformat=parens]{subfig}
\def\BibTeX{{\rm B\kern-.05em{\sc i\kern-.025em b}\kern-.08em
    T\kern-.1667em\lower.7ex\hbox{E}\kern-.125emX}}
\begin{document}

\title{QubiC 2.0: An Extensible Open-Source Qubit Control System Capable of Mid-Circuit Measurement and Feed-Forward}
\author{\IEEEauthorblockN{
Yilun Xu,$^1$
Gang Huang,$^1$
Neelay Fruitwala,$^1$
Neel R. Vora,$^1$
Abhi D. Rajagopala,$^1$
Akel Hashim,$^{1,2}$
Larry Chen,$^2$
Noah Goss,$^2$
Kasra Nowrouzi,$^1$
Ravi K. Naik,$^{1,2}$
Christopher Spitzer,$^1$
David I. Santiago,$^1$
and Irfan Siddiqi$^{1,2}$
}
\IEEEauthorblockA{\\$^1$Lawrence Berkeley National Laboratory, Berkeley, CA 94720, USA
\\$^2$University of California at Berkeley, Berkeley, CA 94720, USA
\\Corresponding author: Gang Huang (email: ghuang@lbl.gov)}}

\maketitle

\begin{abstract}

Researchers manipulate and measure quantum processing units via the classical electronics control system. We developed an open-source FPGA-based quantum bit control system called QubiC for superconducting qubits. After a few years of qubit calibration and testing experience on QubiC 1.0, we recognized the need for mid-circuit measurements and feed-forward capabilities to implement advanced quantum algorithms effectively. Moreover, following the development of RFSoC technology, we upgraded the QubiC system to QubiC 2.0 on an AMD ZCU216 evaluation board and developed all these enriched features.
The system uses portable FPGA gateware with a simplified processor to handle commands on-the-fly. For design simplicity and straightforward scaling, we adopted a multi-core distributed architecture, assigning one processor core per qubit. The actual pulses combine the unique pulse envelope and carrier information specified in a command. Each pulse envelope is pre-stored on FPGA's block RAMs, ensuring the speed and reusability during the whole quantum circuit. The pulse parameters including amplitude, phase, and frequency can be updated from pulse to pulse. The software stack is developed in Python, running on both the FPGA's ARM core and host computer via XML-RPC. The quantum circuit can be described in a high-level language, which supports programming at both the pulse-level and native-gate level, and includes high-level control flow constructs. The QubiC software stack compiles these quantum programs into binary commands that can be loaded into the FPGA. 
With QubiC 2.0, we successfully achieved multi-FPGA synchronization and low-latency data communication, and demonstrated simplified feed-forward experiments on conditional circuits. The system also serves as a platform for artificial intelligence and machine learning driven quantum control, such as FPGA-based real-time quantum state discrimination. The enhanced QubiC system represents a significant step forward in quantum computing, providing researchers with powerful tools to explore and implement advanced quantum algorithms and applications.

\end{abstract}

\section{Introduction}
Quantum computing is a revolutionary field of computing that leverages the principles of quantum mechanics to process information in ways unimaginable by classical computers \cite{preskill2018quantum}. 
By utilizing the quantum bit (qubit), which can exist in multiple states simultaneously, quantum computers have the potential to solve complex problems exponentially faster than classical computers \cite{arute2019quantum}. 
In recent years, quantum computing sparked the development of groundbreaking advances in science \cite{google2023suppressing,kim2023evidence}.
Quantum control, the ability to manipulate and steer quantum systems with precision, lies at the heart of quantum technology advancements \cite{ryan2017hardware,palazzi2022mtt}. 
The field of quantum control has witnessed remarkable progress in recent years \cite{salathe2018low,tholen2022measurement,park2022icarus,stefanazzi2022qick}, driven by a deep understanding of quantum mechanics, novel control techniques, and advancements in experimental capabilities.

As quantum information processors continue to advance, with increasing quantum bit count and complexity, ensuring synchronization among multiple devices becomes crucial for scaling up the system effectively \cite{xu2023fpga}. 
Additionally, the potential to utilize mid-circuit measurement with qubit reset is key to reusing qubits and condensing larger circuits to operate efficiently on smaller quantum devices \cite{hua2022exploiting,decross2022qubit}. 
Mid-circuit measurement plays an important role in quantum error correction, where repeated measurements on ancilla qubits are required \cite{ofek2016extending,botelho2022error}. 
This raises the requirement for a control system to enable a single qubit to be measured at real-time during the execution of the circuit, and further multiple measurements in a circuit and at any point throughout the circuit.

Following the development of the new radio frequency system-on-chip (RFSoC) technique \cite{farley2018all}, we upgraded the QubiC system to QubiC 2.0, fulfilling the need for real-time computation and feedback \cite{huang2023qubic}.
This paper aims to provide an overview of the recent advances in the QubiC 2.0 system and highlight the key methodologies and techniques that have emerged to manipulate and exploit quantum systems. 
We delve into the electronics hardware, field-programmable gate array (FPGA) gateware, engineering software, and synchronization and communication design of QubiC 2.0, examining the strategies employed to achieve real-time control over qubits.
Moreover, we highlight state-of-art fast control techniques, the artificial intelligence and machine learning capabilities integrated into the QubiC ecosystem, and the experimental advancements in mid-circuit measurement and feed-forward over the QubiC 2.0 system.

By exerting fast control over quantum systems, researchers and engineers can harness their unique properties for a wide range of applications, including quantum computing, quantum communication, and quantum sensing.
As an extensible and open source toolbox for the quantum community, QubiC 2.0 is capable of accommodating users and developers from different layers to enable efficient co-design.
Researchers can unlock the full potential of quantum technology, leading to groundbreaking applications and transformative advancements in quantum computing.

\section{Electronics Hardware}

The QubiC system, deployed on the ZCU216 platform, offers advanced radio frequency (RF) data processing and synchronization capabilities. 
With its high-speed digital-to-analog converter (DAC), analog-to-digital converter (ADC) channels, and multi-tile synchronization (MTS), the system provides a powerful and flexible solution for RF data manipulation and adaptive computing. 
Moreover, the integral design of the RFSoC makes the control extensibility cost affordable.  
The QubiC system on the ZCU216 platform represents a significant step towards efficient and seamless RF data processing in quantum computing, opening doors to run advanced real-time experiments on RFSoC system. 

\subsection{ZCU216}
ZCU216 is an evaluation board featuring the Gen 3 ZU49DR Zynq UltraScale+ RFSoC, which provides 930 K system logic cells, 16-ch 14-bit RF-DAC @ 9.85 giga samples per second (GSPS) and 16-ch 14-bit RF-ADC @ 2.5 GSPS \cite{amd2023zcu216}. 
To accommodate ZCU216 to control and measure the qubits, the digital signal processing (DSP) frequency is set to be 500 MHz on ZCU216 in the QubiC system. 
The QubiC 2.0 system integrates a 16-channel 14-bit DAC capable of operating at a high-speed sampling rate of 8 GSPS, and a 2-channel 14-bit ADC with a sampling rate of 2 GSPS, enabling high-fidelity data acquisition.
CLK104 RF clock add-on card is used for internal or external reference clocking, while a quad zSFP/zSFP+ connector is equipped on the ZCU216 board to accept zSFP/zSFP+ modules for data communication between boards.
RFMC 2.0 interfaces are employed as the I/O expansion options for DAC and ADC analog front-end (AFE). 

\subsection{Analog Front-End}

\begin{figure}[t!]
\centering
\includegraphics[width=0.9\linewidth]{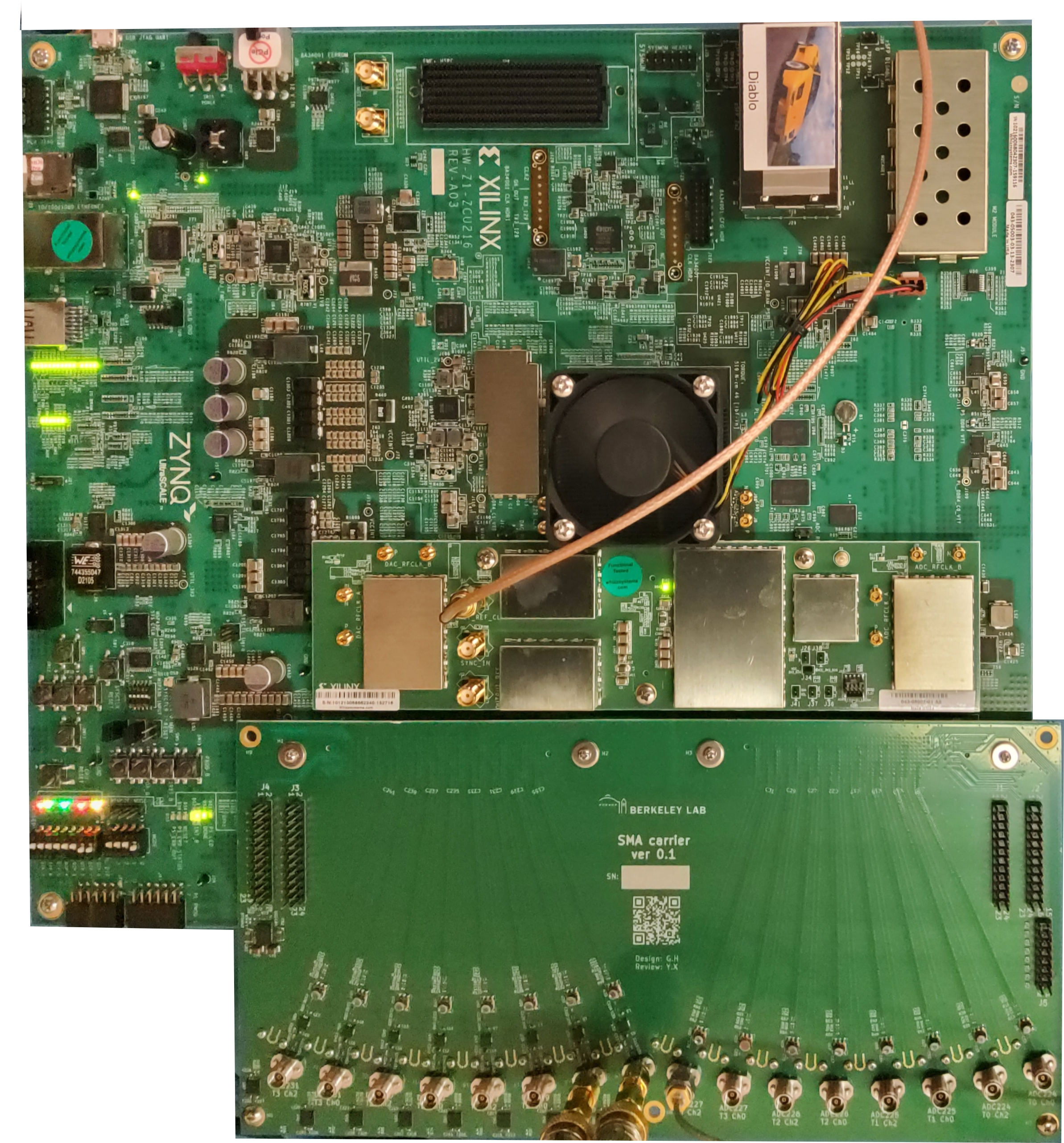}
\caption{Customized analog front-end board and AMD ZCU216 evaluation board.}
\label{fig:afe}
\end{figure}

To fan out / collect signals to / from the fridge, we developed a customized analog front-end board for ZCU216, as shown in Fig.~\ref{fig:afe}. 
This board features two RFMC 2.0 connectors, accommodating 16 DAC and 16 ADC channels respectively. 
A balun is incorporated to convert differential signals to single-ended and vice versa.
To enhance signal quality, the board includes low-noise amplifiers to amplify the DAC signals. 
For power management, we employ switching pre-regulators to ensure power efficiency, while linear low-dropout (LDO) post-regulators provide low-noise power sources.
Moreover, the analog front-end board shares the same physical size as the XM655 breakout add-on card that comes with the ZCU216 kit, making it compatible and interchangeable for enhanced flexibility and ease of integration.
While the analog front-end board shown in Fig.~\ref{fig:afe} is AC-coupled, we recently developed a new DC-coupled analog front-end daughter card that provides flux-tuning capability, extending QubiC 2.0 support to a broader range of superconducting qubit modalities, including transmon, fluxonium, and Kerr-cat qubits.
The detailed design information of the new DC-coupled daughter card is available in the QubiC hardware repository \cite{qubic2025hardware}.

\section{FPGA Gateware} 
Fully open-source QubiC gateware provides the low level access to the researchers, and is capable of generating waveform form parameters, which leads to the benefit of pulse sequence reuse.
The gateware is designed with low latency, and optimized to do the real-time control on FPGA for mid-circuit measurement and feed-forward.
As shown in Fig.~\ref{fig:qubic_gateware}, the QubiC 2.0 FPGA gateware contains a Python Productivity for Zynq (PYNQ) framework on Zynq, advanced extensible interface (AXI) interconnection, a customized intellectual property (IP) for programmable logic (PL) / processing system (PS) units, and RF data converter (RFDC). 
We will mainly discuss the board support package and the digital signal processing (DSP) in the section.

\subsection{Board Support Package}

QubiC 2.0 utilizes the PYNQ 3.0 framework \cite{amd2023pynq} on Zynq UltraScale+ to deliver a user-friendly software platform and a suite of libraries for seamless development of embedded systems and reconfigurable computing applications.
PYNQ streamlines the development process by providing an accessible programming interface and a comprehensive set of libraries tailored for RFSoC-based systems. 
Additionally, it offers a specialized overlay that allows users to effortlessly access and control the RF components, high-speed data converters, and FPGA fabric integrated within the RFSoC device.

The implementation of the AXI, a communication bus protocol integrated on-chip, facilitates interconnection between the Zynq and essential components such as the RFDC, DSP register block, configuration register block, and block random-access memory (BRAM) control block.

The ZCU216 board integrates robust PS and PL units within a single device.
QubiC 2.0 takes advantage of a parameterized IP tailored for the PL and PS, providing the flexibility to accommodate various application requirements effectively.
The DSP registers, configuration registers, and BRAM control blocks are transported to the PLPS block through the local bus, ensuring efficient communication and data access.

\begin{figure}[t!]
\centering
\includegraphics[width=1.0\linewidth]{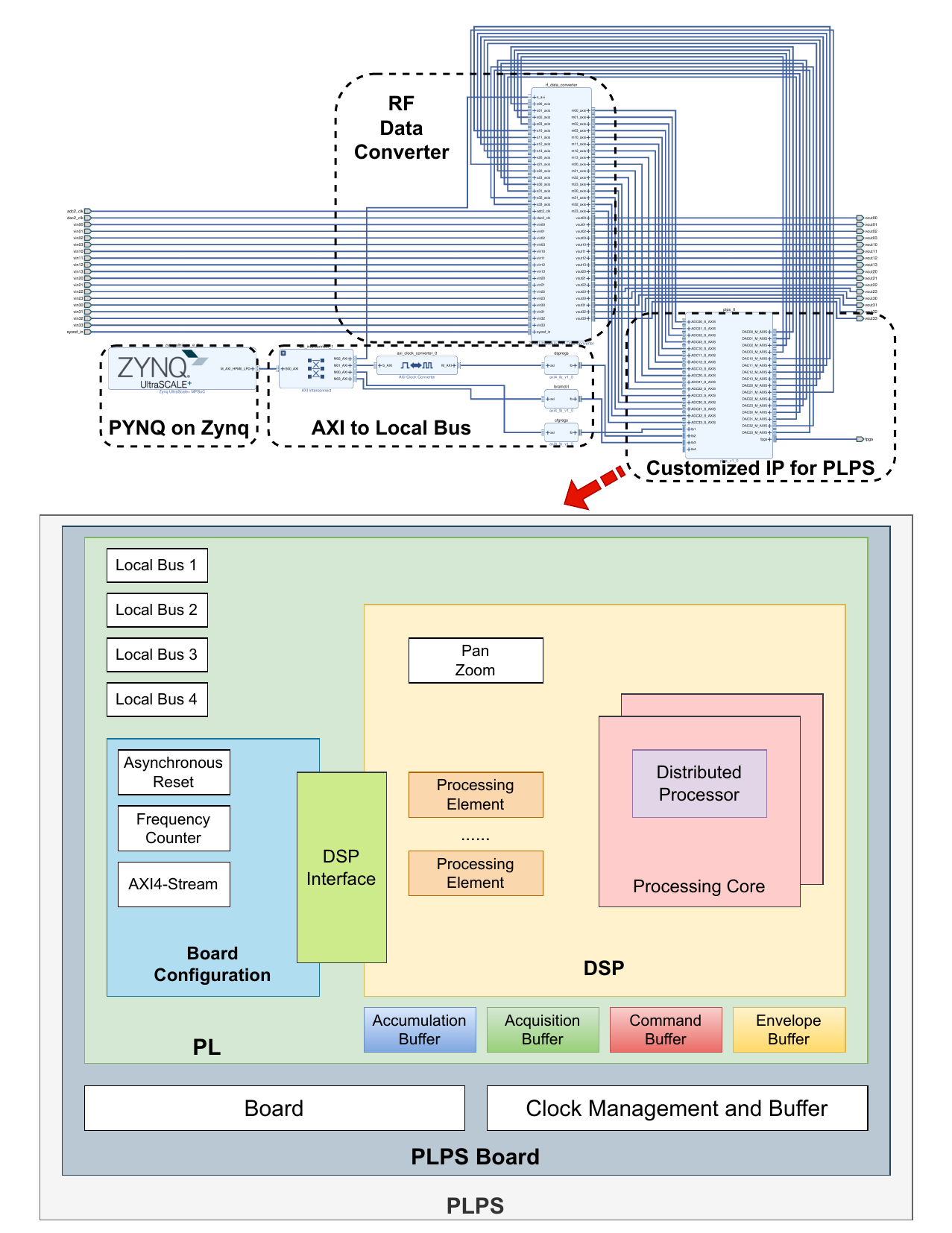}
\caption{QubiC gateware.}
\label{fig:qubic_gateware}
\end{figure}

As shown in Fig.~\ref{fig:qubic_gateware}, in the PLPS block, we integrate programmable logic, clock management / buffer, FPGA pin instantiation, and other hardware connectivity elements.
To facilitate efficient communication between blocks, we employ SystemVerilog \cite{ieee2017sv}, leveraging its interface encapsulation capabilities.
On the PL side, we utilize local buses, board configuration, and the DSP. 
Four local buses are instantiated here for the information exchange with AXI.
The board configuration block comprises the AXI4-Stream interface, meticulously implemented for the data handshake between the DAC/ADC on AXI and the DSP block. 
Additionally, the board configuration block incorporates asynchronous reset logic to minimize latency and a frequency counter, essential for verifying the accuracy of clock frequency generation.
The DSP interface serves as a bridge between the board configuration and the DSP block.

RF data converter is an AMD IP configured with multi-tile synchronization to enable high sampling rate DACs and ADCs. 
To cater to diverse quantum application needs, we offer various variations of RF data converters. 
These variations are carefully designed to strike a balance between qubit drive channels, readout data rate, and FPGA resources, ensuring optimal performance for different quantum computing scenarios.

\subsection{Digital Signal Processing}

The superconducting qubits are precisely controlled and measured using RF pulses, typically amplitude-modulated signals with parameterized frequency, initial phases, and pre-defined envelopes.
To digitally generate the pulses, we employ complex multiplication of the carrier and the modulation envelope. 
Each qubit has its dedicated DAC channel, while the readout drive outputs are combined and directed to a single readout drive DAC, known as multiplexed readout.

The qubit readout signal from the fridge undergoes digitization by ADC. 
Next, it is mixed with a digital local oscillator, and the resulting signal is integrated during the duration of the readout pulse to obtain the IQ data.
The accumulation process acts as a low-pass filter, down-converting the signal to the frequency of interest, and the filtered data is then stored in the corresponding buffer for subsequent qubit state discrimination.

\begin{figure}[t!]
\centering
\includegraphics[width=1.0\linewidth]{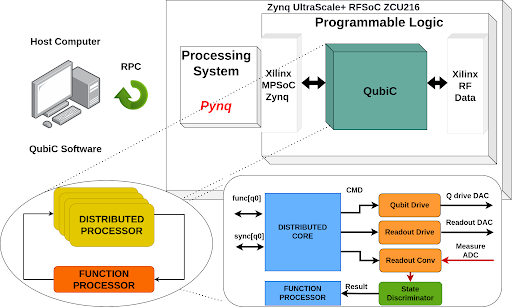}
\caption{QubiC architecture.}
\label{fig:qubic_arch}
\end{figure}

There is a bank of soft-processor cores capable of parameterizing and triggering RF control/measurement pulses synthesized using direct digital synthesis (DDS)-based pulse generators on the FPGA gateware DSP.
The cores are simple and lightweight, so each core is intended to only connect to a small number of signal generators.
Since the processor is distributed in nature, with one core per arbitrary waveform generator (AWG), simplifying the design and allowing for straightforward scaling, we call it the distributed processor \cite{fruitwala2022distributed}, as shown in Fig.~\ref{fig:qubic_arch}.
Feedback is implemented at the lowest level of the classical electronics to give highest bandwidth.  

The distributed processor consists of a set of cores, along with interfaces for receiving processed quantum measurement results and synchronization among different cores.
Each processor core executes a program consisting of timed pulse commands for its AWG(s), as well as instructions for implementing arbitrary control flow within the program (e.g. conditional branching) based on either external input or internal state. 
For the timed pulse commands, we largely use the same design as the existing QubiC command buffer \cite{xu2021qubic}: a 72-bit command specifying pulse characteristics (such frequency, phase, length, and pointer to a buffer containing the envelope shape) is sent to the AWG at a specified time, referenced to an internal counter. 
To implement control flow instructions, we include a 16x 32-bit register bank along with an arithmetic logic unit (ALU) that implements comparisons, addition, and subtraction on signed 32-bit integers. 
ALU outputs can be used to write to registers, increment the time reference counter, and set the instruction pointer. 
In the current version of the processor (running at 500 MHz), pulse commands can execute in $\geq 4$ clock cycles, and most register writes/control flow instructions execute in 4-7 clock cycles. 

In each processor core, we include a function processor interface for connecting it to external computational resources. 
This external result could be a state-discriminated measurement, or a more complex computation, such as the output of a decoder for an error correction algorithm. 
The interface is triggered by a special instruction, which halts the execution inside the core until the result returns, at which point the data is either read into a register or used immediately as the input to a branch conditional.

A synchronization interface, along with a special instruction, is employed to allow for the internal time references from different cores to be synchronized. 
Once all of the cores referenced have reached the corresponding sync instruction, program execution within all cores is resumed, and each core's internal time reference is reset.

The processing elements play a pivotal role as up or down converters in the digital domain, efficiently bridging the DSP interface with the distributed processors.
Within the DSP, we incorporate the command buffer and the envelope buffer to facilitate parameterized pulse generation.
Furthermore, the accumulation buffer stores the accumulated values resulting from the integration of the baseband I/Q series over the corresponding digital local oscillator (DLO).
Additionally, the acquisition buffer functions as a live oscilloscope, enabling real-time observation of the ADC/DLO/DAC raw data, enhancing the system's versatility and usability. 

Notably, the QubiC 2.0 gateware DSP empowers the parameterized pulse generation feature.
It enables the efficient reuse of sequences by updating waveform parameters through commands without the need for recompiling or resending all the waveform data. 
The pulse envelope is carefully optimized during the qubit gate calibration process and is stored in the FPGA memory as complex numbers, with each point containing in-phase and quadrature-phase terms.  
This stored pulse envelope can be conveniently reused by indexing the same address area, even if there are changes in the carrier frequency or phase.
Running a quantum circuit on QubiC 2.0 is remarkably efficient as it loads only the required commands instead of the entire waveform data points, streamlining the computational process and maximizing system performance.

\section{Engineering Software}
\begin{figure}[t!]
\centering
\includegraphics[width=1.0\linewidth]{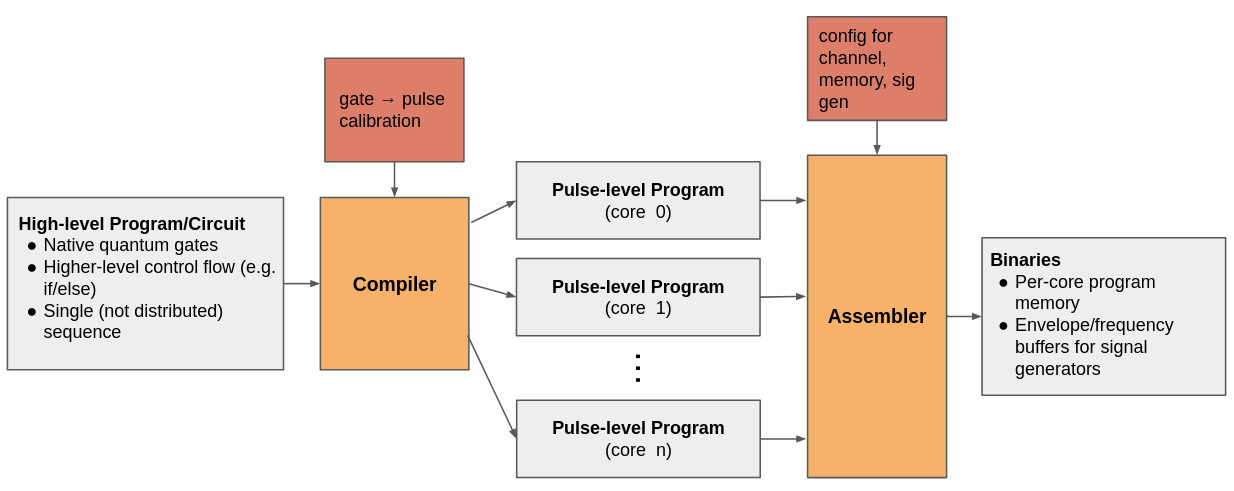}
\caption{QubiC software stack.}
\label{fig:software}
\end{figure}

The low level toolchain contains the compiler and the assembler, as shown in Fig.~\ref{fig:software}. 
The input to the compiler is a high level program or circuit description as a sequence of gates and/or pulses together with high level control flow (e.g. if/else for branching) structures. 
The compiler converts this description into a custom intermediate representation (IR), then implements a series of passes to resolve gates into pulses, apply virtual phases, schedule pulses. 
After all IR passes are complete, the compiler generates a pulse level program for each distributed processor core. 
The current QubiC system uses a mapping of one core per qubit, but this configuration is arbitrary and can be specified as an option to the compiler.
The assembler converts each pulse level program to program binaries for the distributed processor to execute on the FPGA. 
The hardware configuration for each channel is used in the assembling process to determine channel mapping and pulse-generator parameters.

The compiler and assembler can be executed either on the advanced RISC machines (ARM) core on the Zynq for the simple system configuration or on a host computer for the more advanced computational power compared to the embedded core. 
A simple XML-RPC is used for the host-embedded core communication. 
Users can execute the code on a Jupyter Notebook hosted on the host computer, control and read the results by read and write the FPGA buffers. 

\section{Synchronization and Communication}
The number of channels can easily exceed the capability of a single FPGA or RFSoC to handle, making synchronization among multiple devices vital for scaling up the system effectively.
Regarding clock synchronization, the primary objective is to achieve precise phase alignment for all clock signals in each module. This ensures that the DAC and ADC signals are perfectly synchronized in phase, and the clock counter is accurately aligned \cite{huang2021clock}.
Beyond clock synchronization, mid-circuit measurement and feed-forward across boards further require low-latency data communication to distribute quantum measurement results among multiple FPGAs \cite{xu2025multifpga}.

\subsection{Deterministic Clock Distribution}
To achieve synchronization, the clock must be distributed to each node through matched-length cables, ensuring aligned and deterministic phase relationships.
Nested dual phase-locked loop (PLL) in zero-delay mode (ZDM) is employed in LMK04828 of ZCU216 for deterministic clock distribution \cite{ti2023lmk04828}. 
The PLLs' input and feedback frequencies are set to the greatest common divisor among all frequencies, and SYSREF serves as the feedback signal.
It is essential to implement divider reset after the ZDM configuration, as this step is critical for the proper functioning of the clock synchronization system.

\subsection{Multi-DAC Synchronization on Single FPGA}
Multi-tile synchronization (MTS) is a vital feature to synchronize operation across multiple tiles, and facilitate seamless coordination and data transfer between different DAC and ADC elements \cite{amd2023mts}.
To achieve deterministic latency, MTS utilizes an external SYSREF signal distributed to each device as a trigger, aligning frame and local multiple frame clocks to it.
To perform MTS on ZCU216, the first step involves enabling all clocks and SYSREF generators, followed by SYSREF analog capture. 
Subsequently, the clock divider must be reset, and FIFO latency should be measured and adjusted accordingly. 
Finally, digital features are synchronized with SYSREF dynamic update events to complete the MTS setup. 

\subsection{Multi-FPGA Synchronization}
To achieve multi-FPGA synchronization, the FPGAs should share an identical reference clock. 
Following the deterministic clock distribution process, we proceed to implement the simplified precision time protocol \cite{ieee2019ptp}. 

\begin{figure}[t!]
\centering
\includegraphics[width=0.5\linewidth]{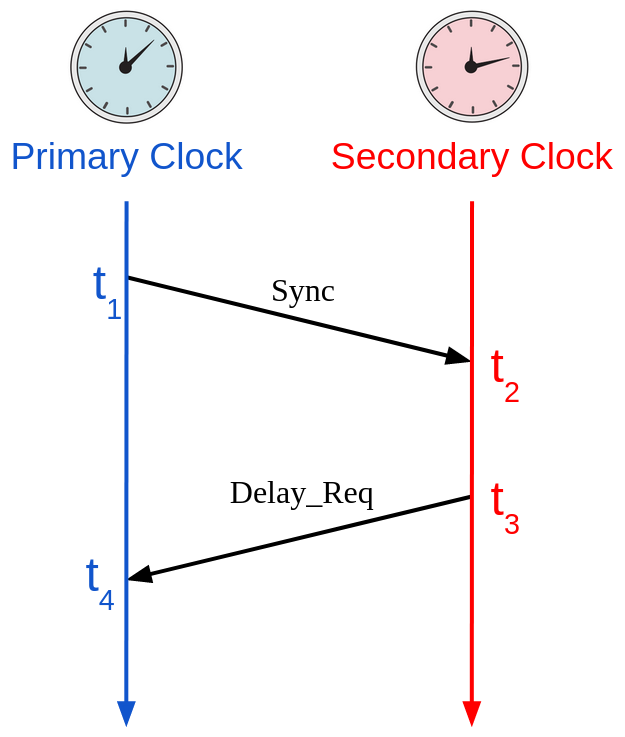}
\caption{Simplified precision time protocol.}
\label{fig:ptp}
\end{figure}
 
As illustrated in Fig.~\ref{fig:ptp}, to synchronize the primary and secondary clocks on two FPGAs, a \textit{Sync} message is transmitted at $t_1$ in the primary clock domain, and the secondary FPGA receives it at $t_2$ in the secondary clock domain. 
Subsequently, the secondary FPGA responds with a \textit{Delay\_Req} message at $t_3$ in the secondary clock domain, and the primary FPGA receives it at $t_4$ in the primary clock domain.
The transit time for the \textit{Sync} message can be calculated as $delay=\frac{(t_4-t_1)-(t_3-t_2)}{2}$, while the constant offset between the primary and secondary clocks can be determined as $offset=\frac{(t_2-t_1)-(t_4-t_3)}{2}$.
To synchronize the secondary clock with the primary clock, one can apply the offset correction in software, ensuring precise synchronization between the two clock domains.

\subsection{Data Communication}
Real-time feed-forward across boards requires the measurement results to be distributed with a latency that is only a small fraction of the qubit coherence time.
We built the data communication over fiber directly on the raw GTY gigabit transceivers with a minimal custom protocol tailored to quantum measurement data \cite{xu2025multifpga}.
The design utilizes all four small form-factor pluggable (SFP28) interfaces on the ZCU216, each providing a payload rate of 25 Gbps.
The transmitter continuously streams a scrambled 64-bit measurement word with no packets, idle sequences, checksums, or handshakes on the link, so the loss of any individual block is self-healing and the end-to-end latency is deterministic at the clock-cycle level.
The data communication module integrates directly with the distributed processor through its function processor interface: each processor core can read any remote measurement result, or branch on it, using the same instructions as for local measurements, making multi-board feed-forward transparent to the compiled quantum program.

\subsection{Experimental Results}
\begin{figure}[t!]
\centering
\includegraphics[width=0.8\linewidth]{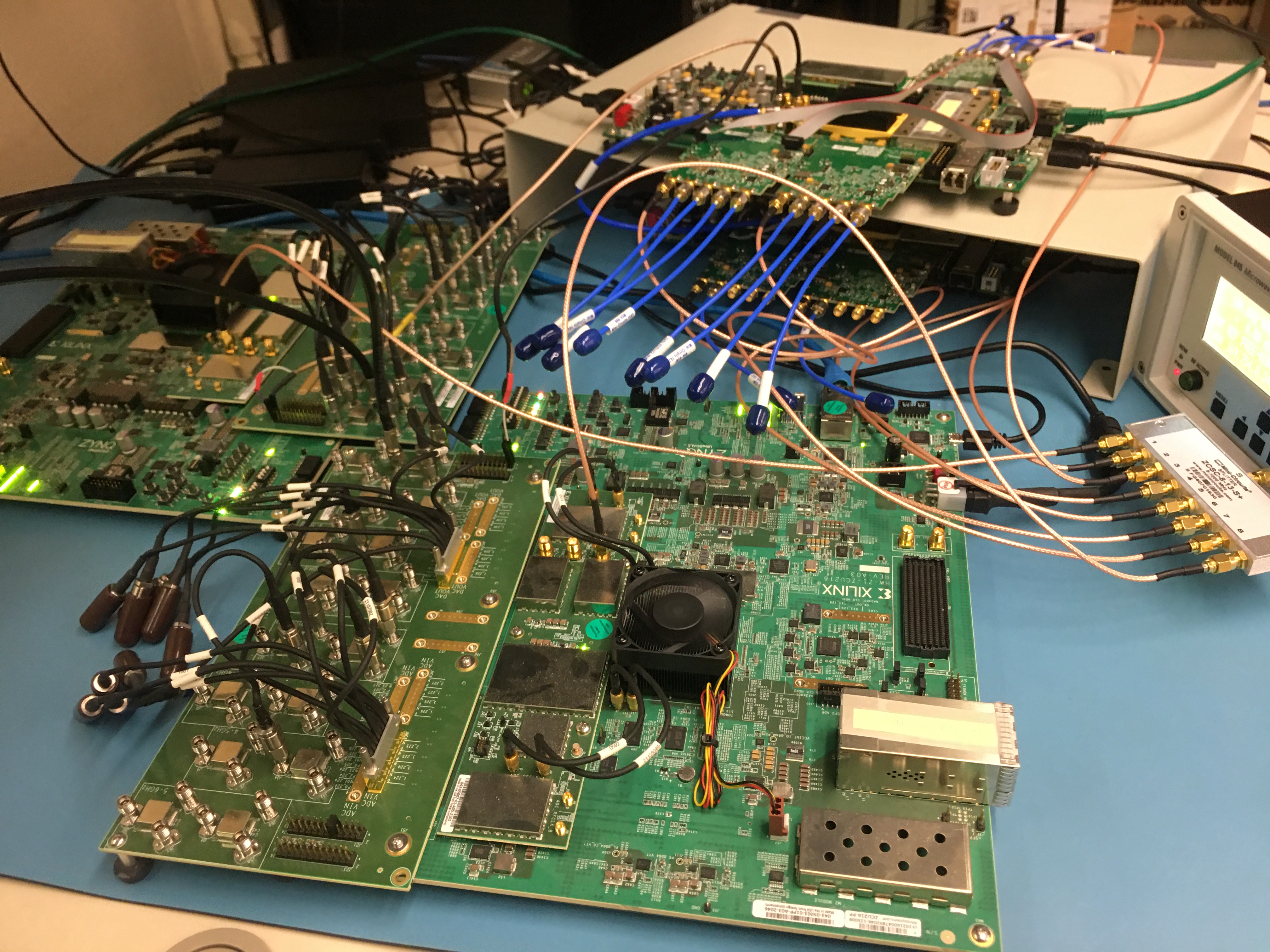}
\caption{Multi-FPGA synchronization hardware.}
\label{fig:sync_hw}
\end{figure}

We configured the multi-FPGA hardware setup, as depicted in Fig.~\ref{fig:sync_hw}, and validated both the clock synchronization and the data communication through bench tests and qubit experiments \cite{xu2025multifpga}.
The signal root mean square (RMS) jitter @ 6.50 GHz on ZCU216 was measured to be 1.8 ps (integrating from 10 Hz to 100 MHz) and 7.4 ps (integrating from 0.1 Hz to 100 MHz), which effectively meets the requirements for quantum experiments, and the clock counters across boards remained aligned with zero offset over a 16-hour test.
Ramsey and Hahn echo coherence experiments driven by two clock-synchronized boards show no measurable degradation compared to single-board operation.
The data communication link achieves a one-way board-to-board latency of 71.2 $\pm$ 2.6 ns and enables a real-time cross-board conditional bit-flip experiment, demonstrating mid-circuit measurement and feed-forward across multiple boards well within the qubit coherence time.
The detailed experimental results are reported in \cite{xu2025multifpga}.

\section{Artificial Intelligence and Machine Learning}
The complexity inherent in quantum mechanical systems renders them particularly apt for high-dimensional, data-driven learning models, and interfacing quantum information science with artificial intelligence (AI) and machine learning (ML) opens up possibilities for more robust and scalable quantum computation and experimental techniques.
Traditional control approaches can be cumbersome or inaccurate for high-dimensional or noise-sensitive systems, whereas AI-driven control provides versatile tools for optimizing and adapting control strategies when analytical solutions are impractical, along with automation capability that reduces both cost and time.
Therefore, AI/ML techniques are being integrated into the QubiC ecosystem to enable AI-driven quantum control \cite{xu2025artificial}.
We demonstrated the first ML-powered quantum state discrimination deployed on an RFSoC FPGA, achieving 66 ns inference latency and 97\% qutrit classification fidelity with a readout pulse as short as 500 ns, enabling mid-circuit measurements \cite{vora2024ml}.
Complementarily, time-series-based quantum state discrimination with long short-term memory (LSTM) networks operates on raw readout trajectories to capture transient decay events missed by conventional methods, yielding consistent fidelity improvements \cite{jung2026time}.
Ongoing developments include online learning on FPGA for adaptive readout correction, where backpropagation is implemented directly on the FPGA to adapt in real time to drifting readout distributions without transferring data to external compute; reinforcement learning for autonomous gate calibration, where a soft actor-critic agent combined with robust phase estimation fine-tunes single-qubit gates and demonstrates coherent error reduction across multiple cooldowns without full recalibration; and agentic AI for quantum calibration, where a large language model powered agent autonomously orchestrates qubit and gate calibration workflows, interprets results, diagnoses failures, and sequences tasks to minimize human intervention \cite{xu2025artificial}.
Through the LBNL--NVIDIA NVQLink partnership, QubiC has been integrated with GPU-accelerated computing to establish real-time feedback loops between the quantum processing unit and the GPU \cite{caldwell2025platform}.
In addition, an open-source data storage and visualization platform for collaborative qubit control provides the training data for these AI/ML workflows \cite{brahmbhatt2024open}.

\section{Demonstration on Qubits}
\subsection{Fast Reset}
Fast reset is a technique utilized to swiftly and efficiently restore a quantum system to a well-defined initial state, commonly the ground state \cite{magnard2018fast}.
This process of resetting qubits plays a vital role in quantum computing, enabling researchers and engineers to prepare quantum states and execute quantum algorithms with greater efficiency.
Conventional quantum reset procedures are often time-consuming and involve dissipative methods that result in the loss of coherence and valuable quantum information.
In contrast, fast reset techniques aim to overcome these limitations by employing non-dissipative and expedited methods to reset qubits while preserving their coherence and quantum properties.
 
Fast reset is a fundamental qubit control feedback loop, where low hardware latency is of utmost importance.
On the QubiC 2.0 system, we developed fast reset logic by implementing qubit status classification and the feedback loop within the FPGA gateware, ensuring minimal latency.
The method involves a single-shot measurement of the qubit state, followed by a conditional single-qubit gate operation. 
If the qubit is found in an excited state, this gate operation rotates it into the ground state.
To enhance the effectiveness of qubit reset, the feedback cycle can be repeated as needed, ensuring the qubits are in a well-defined initial state for subsequent quantum computations.

\begin{figure}[t!]
\centering
\subfloat[Fast reset circuit.]{\includegraphics[width=0.8\linewidth]{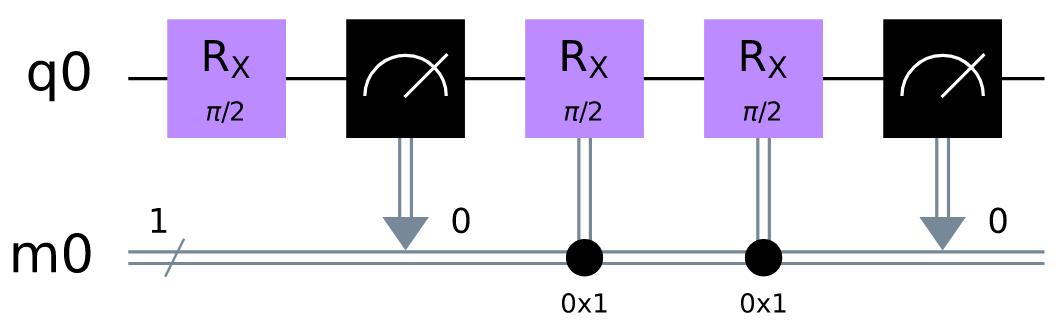}
\label{fig:fastreset_circ}}
\vfil
\subfloat[Fast reset measurement result.]{\includegraphics[width=0.8\linewidth]{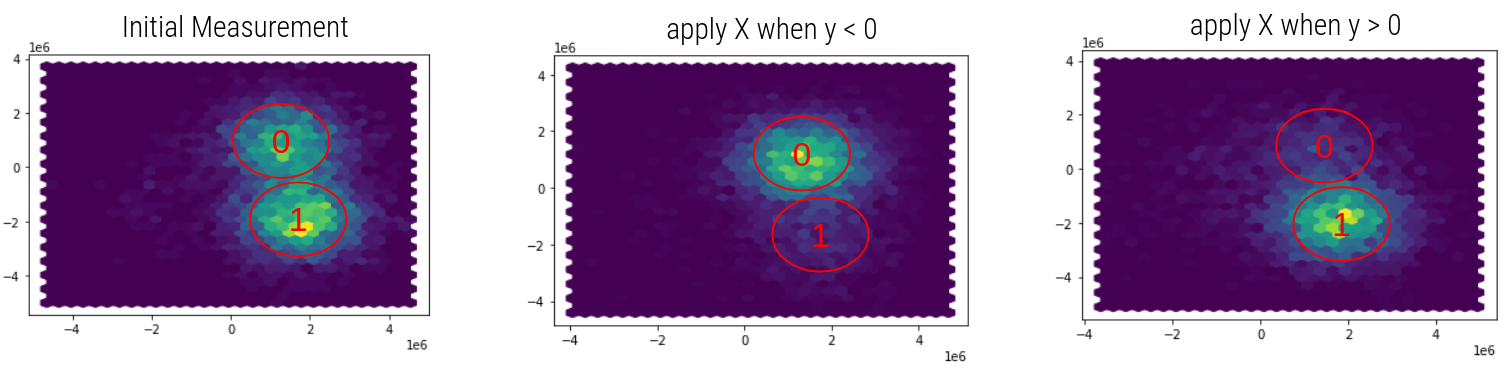}
\label{fig:fastreset_result}}
\caption{Fast reset.}
\label{fig:fastreset}
\end{figure}

As illustrated in Fig.~\ref{fig:fastreset_circ}, the qubit is initialized with an X90 gate, positioning it at the equator of the sphere.
The blobs on the IQ plane are shifted and rotated to create a decision boundary along the X-axis. 
The states $|0 \rangle$ and $|1 \rangle$ are initially equally weighted in the measurement.
In the left plot of Fig.~\ref{fig:fastreset_result}, the decision criterion for state $|0 \rangle$ is $y>0$, whereas state $|1 \rangle$ is determined by $y<0$.
To test the functionality of fast reset, when the mid-measurement state is $|1 \rangle$, we apply two consecutive X90 gates, rotating the qubit state back to $|0 \rangle$. 
As shown in the middle plot of Fig.~\ref{fig:fastreset_result}, state $|0 \rangle$ is observed, aligning with our expectations.
Conversely, when the mid-measurement state is $|0 \rangle$, we apply two consecutive X90 gates, rotating the qubit state to $|1 \rangle$.
As illustrated in the right plot of Fig.~\ref{fig:fastreset_result}, state $|1 \rangle$ is observed as anticipated, further validating the effectiveness of the fast reset process.
Preliminary experiments reveal significant reductions in experiment run-time due to the active reset, all while maintaining accuracy and preserving quantum coherence.

\subsection{Conditional Bit-flip}
Mid-circuit measurement involves performing measurements on specific qubits at designated times during the execution of a quantum circuit. 
Feed-forward is a technique that enables the implementation of conditional operations based on the outcomes of previous quantum measurements.
Mid-circuit measurement and feed-forward experiment hold significant implications for quantum algorithms and computations, including state projection, branching, measurement-based quantum computing, and error correction.

With QubiC 2.0's capabilities, mid-circuit measurement and feed-forward experiments are achievable, enabling the realization of advanced quantum algorithms. 
As an example, a simplified conditional bit-flip circuit is demonstrated on two qubits \cite{fruitwala2023distributed}.

\begin{figure}[t!]
\centering
\subfloat[Conditional bit-flip circuit.]{\includegraphics[width=0.8\linewidth]{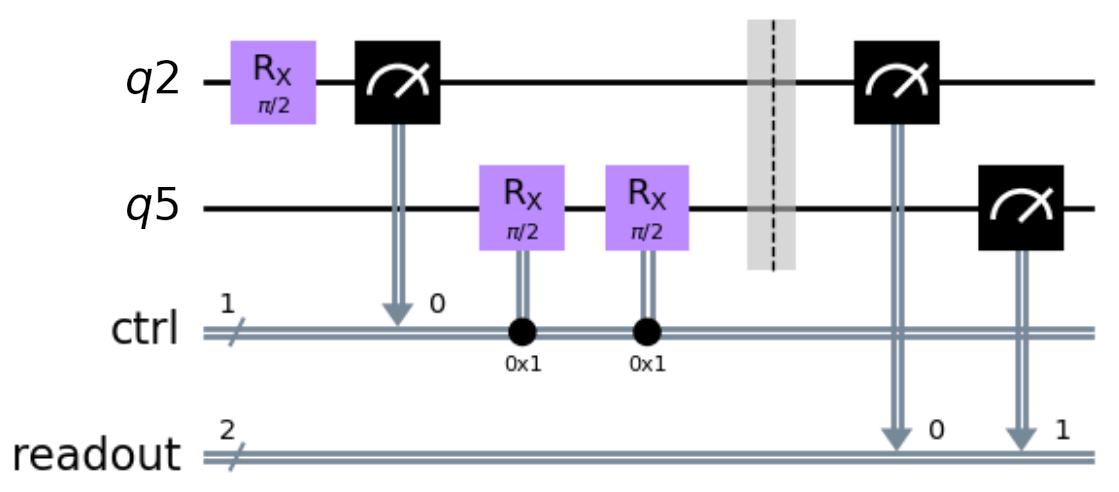}
\label{fig:condbitflip_circ}}
\vfil
\subfloat[Conditional bit-flip result.]{\includegraphics[width=0.8\linewidth]{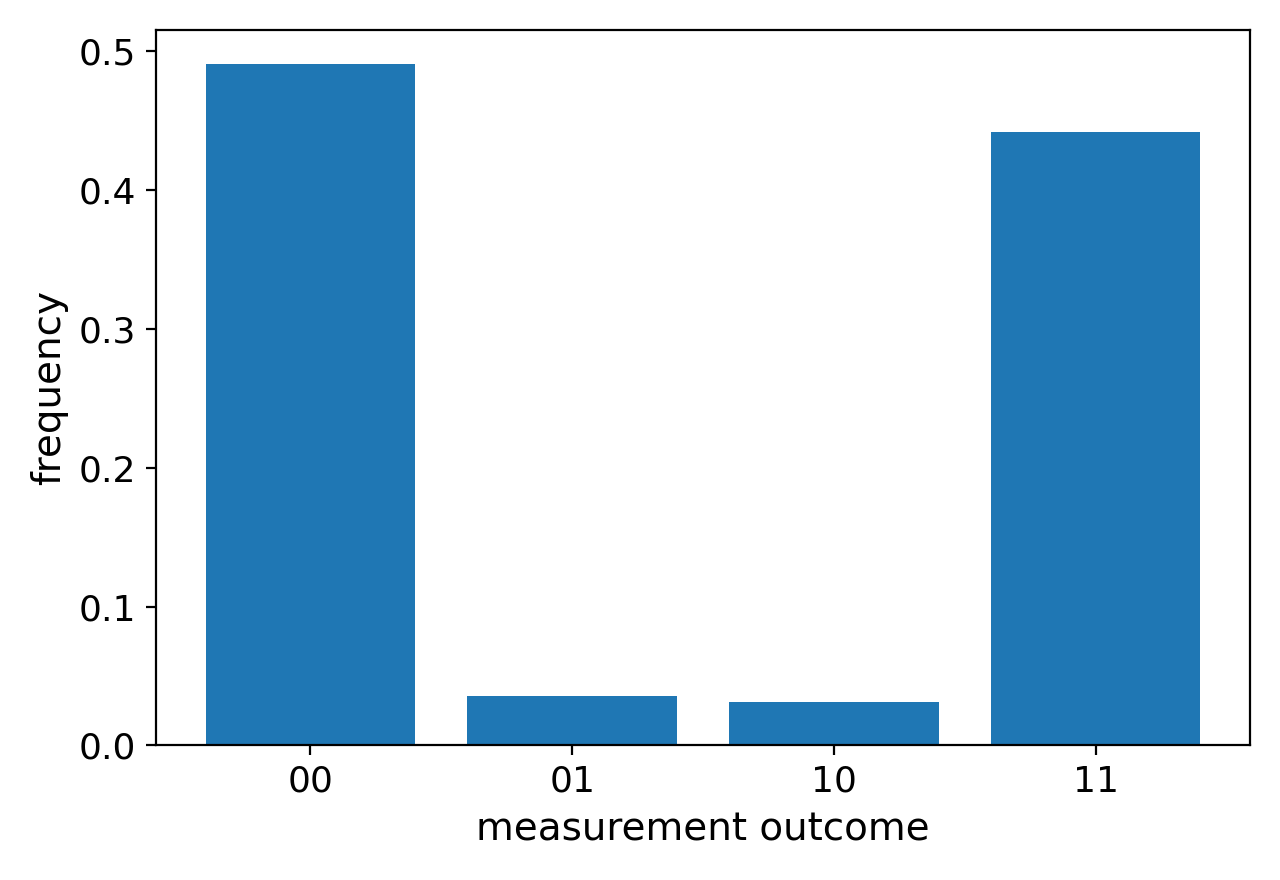}
\label{fig:condbitflip_result}}
\caption{Conditional bit-flip.}
\label{fig:condbitflip}
\end{figure}

As depicted in Fig.~\ref{fig:condbitflip_circ}, the control qubit (Q2) is initialized to a state on the equator, and a mid-circuit measurement is performed to condition the gate operation on the target qubit (Q5).
If the Q2 mid-circuit measurement yields $|1 \rangle$, two consecutive X90 gates will be applied on Q5.
Conversely, if Q2 is measured to be $|0 \rangle$, no operation is performed on Q5.
The experiment was executed with 10,000 shots without any readout error mitigation.
As observed in Fig.~\ref{fig:condbitflip_result}, the correlated outcomes $|00 \rangle$ and $|11 \rangle$ dominate the final measurement with balanced probabilities of 0.491 and 0.442.
The residual $|01 \rangle$ and $|10 \rangle$ populations of 0.036 and 0.032 are consistent with the independently characterized readout fidelities of the two qubits ($P(0|0)=0.984$ and $P(1|1)=0.959$ for Q2, and $P(0|0)=0.977$ and $P(1|1)=0.966$ for Q5), showcasing the successful mid-circuit measurement and feed-forward functionality of the system.

\section{Conclusion}
We developed and upgraded the QubiC 2.0, which is an extensible open-source qubit control system capable of mid-circuit measurement and feed-forward. 
The system was built on the AMD RFSoC ZCU216 together with customized AFE boards, including a new DC-coupled daughter card for flux tuning that supports transmon, fluxonium, and Kerr-cat qubits.
A portable FPGA gateware with distributed processors is implemented to handle the commands on-the-fly. 
QubiC 2.0 gateware enables the parameterized pulse generation, and reuses sequences by updating waveform parameters via commands without recompiling or resending all the waveform data. 
The full software stack is developed using Python on top of the PYNQ platform, running on the
FPGA’s ARM core, while an XML-RPC protocol is adopted to communicate between the host computer and the ARM core. 
The quantum circuit can be described in the high level language with gate and/or pulse information, and got compiled/assembled to the binary commands which are loaded to the FPGA.
We successfully achieved multi-FPGA synchronization and low-latency data communication, and demonstrated the fast reset and conditional bit-flip experiments on superconducting qubits.
Furthermore, the QubiC ecosystem integrates artificial intelligence and machine learning capabilities, ranging from real-time quantum state discrimination on FPGA to autonomous calibration, and has been coupled with GPU-accelerated computing for real-time feedback.
The extensible QubiC system with mid-circuit measurement and feed-forward capabilities provides researchers with powerful tools to explore and implement advanced quantum algorithms and applications.

\section*{Acknowledgment}
This work was supported by the U.S. Department of Energy, Office of Science, Advanced Scientific Computing Research Testbeds for Science program, the National Quantum Information Science Research Centers Quantum Systems Accelerator, the High Energy Physics QUANTISED program, and the Laboratory Directed Research and Development (LDRD) funding from the Lawrence Berkeley National Laboratory under Contract No. DE-AC02-05CH11231.

Y.X. has a financial interest in OpenQI, Inc.

\bibliographystyle{IEEEtran}
\bibliography{reference}

\end{document}